\documentclass{article}
\usepackage{graphicx}
\usepackage{float}
\usepackage[final]{nips_2017}

\usepackage[utf8]{inputenc} 
\usepackage[T1]{fontenc}    
\usepackage{hyperref}       
\usepackage{url}            
\usepackage{booktabs}       
\usepackage{amsfonts}       
\usepackage{nicefrac}       
\usepackage{microtype}      

\title{Neural Style Transfer for Audio Spectrograms }

\author{
Prateek Verma, Julius O. Smith \\
Center for Computer Research in Music and Acoustics (CCRMA), Stanford University\\
\texttt{prateekv@stanford.edu, jos@ccrma.stanford.edu}}
 
\begin{document}

\maketitle
%
\begin{abstract}
There has been fascinating work on creating artistic transformations of images by Gatys et al. This was revolutionary in how we can in some sense alter the ``style'' of an image while generally preserving its ``content''.  In this work, we present a method for creating new sounds using a similar approach. For demonstration, we investigate two different tasks, resulting in bandwidth expansion/compression, and timbral transfer from singing voice to musical instruments. 
\end{abstract}

\section{Introduction}
We present a new machine learning technique for generating music and audio signals. The focus of this work is to develop new techniques parallel to what has been proposed for artistic style transfer for images by Gatys et al.\ [1].  We present two cases of modifying an audio signal to generate new sounds. A feature of our method is that a single architecture can generate these different audio-style-transfer types using the same set of parameters which otherwise require complex hand-tuned diverse signal processing pipelines. Finally, we propose and investigate generation of spectrograms from noise by satisfying an optimization criterion derived from features derived from filter-activations of a convolutional neural net. The potential flexibility of this sound-generating approach is discussed.

\section{Methodology} 	

There has been recent work applying architectures in computer vision for acoustic
scene analysis.  In particular, [3] uses standard architectures such
as AlexNet, VGG-Net, and ResNet for sound understanding. The
performance gains from the vision models are translated to the audio
domain as well. The work in [2, 3] used a mel-filter-bank input
representation, while we use Short-Time Fourier Transform (STFT)
log-magnitude instead.  We desire a high-resolution audio
representation from which perfect reconstruction is possible via,
e.g., Griffin-Lim Reconstruction [7]. All experiments in this work use
an audio spectrogram representation having duration 2.57s, frame-size
30ms, frame-step 10ms, FFT-size 512, and audio sampling rate of 16kHz.

The core of the success of neural style transfer for vision is to
optimize the input signal, starting with random noise, to take on the
features of interest derived from activations at different layers
after the passing through a convolutional net based classifier which
was trained on the content of the input image. We follow a similar approach,
with some modifications for audio signals. First, we train a standard AlexNet [5]
architecture, but have a smaller receptive size of $3 \times 3$
instead of the larger receptive fields used in the original work. This
is to retain the audio resolution, both along time and frequency, as
larger receptive fields would yield poor localization in the audio
reconstruction, which results in audible artifacts. We also add
additional loss terms in order to match the averaged timbral and
energy envelope. All applications here correspond to \emph{timbre
  transfer} of musical instruments having no explicit knowledge of
features such as pitch, note onset time, type of instrument, and so
on. The AlexNet was trained on audio spectograms to distinguish classes of musical
instrument sounds ($80$ from AudioSet), with $3 \times 3$ convolutions
and $2\times 2$ pooling, having a total of $6$ layers with objective
function minimizing the cross-entropy loss using the Adam optimizer
[4].

\section{Experiments}

 We focus on two experiments: (1) imposing the style of a tuning fork
 on a harp, resulting in bandwidth compression down to the
 fundamental, and (2) transferring the style of a violin note to a
 singing voice, resulting in bandwidth expansion. Thus, we have a new
 form of cross-synthesis imposing the style of one instrument on the
 content of another, with applications similar to [6]. We explored
 various hyper-parameters and single/multiple layers from which we extract these
 features for optimization. The goal is to have a single parameter
 setting that can perform all of these tasks, without having to
 explicitly develop hand-crafted rules. Traditionally there have been
 distinct signal processing based approaches to do such
 tasks. Subplots in Figs.{} 1-2 a)-d) are log-magnitude spectrograms
 with the y-axis 0-8kHz and x-axis 0-2.57s. Note in Fig. 2. how this
 approach not only changes the timbre, but also increases the
 bandwidth of the signal, as seen in the strength of the higher
 harmonics. The objective equation below drives the reconstructed
 spectrogram $X_{recon}$ from random noise to be the spectra that minimizes the sum of
 weighted loss terms $L_c$ denoting the \emph{content loss} (the
 Euclidean norm of the difference between the current activation
 filters and those of the content spectrogram), $L_s$ the \emph{style
   loss} (which is the normalized Euclidean norm between the Gram
 matrix of filter activations of selected convolutional layers similar
 to [1] between $X$ and $X_s$), and $L_e$ and $L_t$ which measure
 deviation in the temporal and frequency energy envelopes
 respectively from the style audio. We found that matching the weighted energy contour and
 frequency energy contour (timbral envelope), namely $e_s$ and $t_s$, averaged over time in
 our loss function, helped in achieving improved quality. The
 energy term in the loss function is required because the Gram matrix
 does not incorporate temporal dynamics of the target audio style,
 and would generally follow that of the content if not included.%

\begin{eqnarray*}
X_{recon}&=&\mbox{argmin}_X\mathcal {L}_{\mbox{total}}
=\mbox{argmin}_X\; \alpha L_c(x,x_c)+\beta L_s(x,x_s)+\gamma L_e(x_e,e_s)+\delta L_t(x_t,t_s).  
\end{eqnarray*}
 
\vspace{-1.5em}
 
 \begin{minipage}{\linewidth}
      \centering
      \begin{minipage}{.45\linewidth}
          \begin{figure}[H]
              \includegraphics[width=2in, height=1.8in]{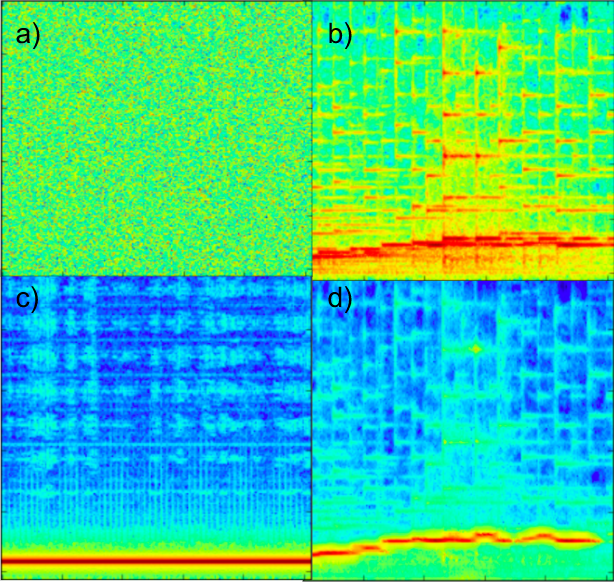}
              \caption{a) shows the Gaussian noise from which we start the input to optimize, b) Harp sound (content) c) Tuning Fork (style) and d) Neural Style transferred output with having content of harp and style of tuning fork \url{https://youtu.be/UlwBsEigcdE}
}
          \end{figure}
      \end{minipage}
      \hspace{0.05\linewidth}
      \begin{minipage}{.45\linewidth}
          \begin{figure}[H]
              \includegraphics[width=2in,height=1.8in]{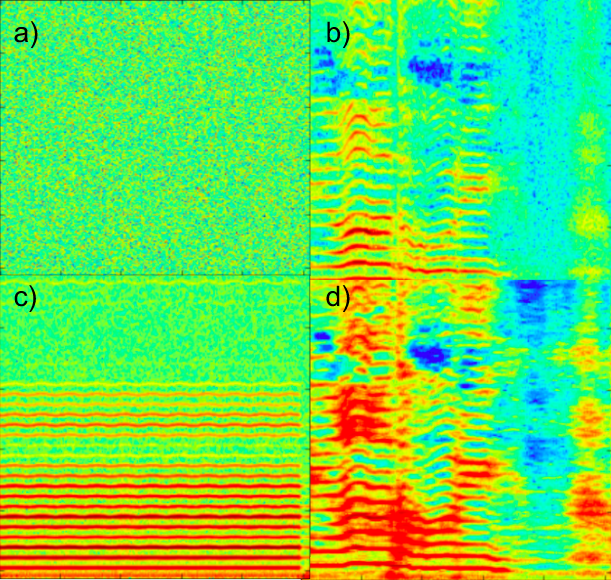}
              \caption{a) shows the Gaussian noise from which we start the input to optimize, b) Singing sound (content) c) Violin note (style) and d) Neural Style transferred output with having content of singing and style of violin. \url{https://youtu.be/RpGBkfs24uc}}
          \end{figure}
      \end{minipage}
  \end{minipage}

\section{Conclusion and Future Work}
			
We have proposed a novel way to synthesize audio by treating it as a
style-transfer problem, starting from a random-noise input signal and
iteratively using back-propagation to optimize the sound to conform to
filter-outputs from a pre-trained neural architecture.  The two
examples were intended to explore and illustrate the nature of the
style transfer for spectrograms, and more musical examples are
subjects of ongoing work.  The flexibility of this approach, and the
promising results to date indicate interesting future sound
cross-synthesis methods. We believe this work can be extended to
many new audio synthesis/modification techniques based on new
loss-term formulations for the problem of interest, and are excited to see and hear what lies
ahead.
\footnote{Acknowledgments: The authors would like to thank Andrew Ng's group and the Stanford Artificial Intelligence Laboratory for the use of their computing resources. Prateek Verma would like to thank Ziang Xie for discussion about challenges in the problem, and Alexandre Alahi for style transfer work in computer vision.}

\noindent
{\Large \bf References}\\
\begin{enumerate}
\item [1.] Gatys, Leon A., Alexander S. Ecker, and Matthias Bethge. "A neural algorithm of artistic style." arXiv preprint arXiv:1508.06576(2015).
\item [2.] Recommending music on Spotify with deep learning – Sander Dieleman benanne.github.io/2014/08/05/spotify-cnns.html
\item [3.] Hershey, Shawn, et al. "CNN architectures for large-scale audio classification." Acoustics, Speech and Signal Processing (ICASSP), 2017 IEEE International Conference on. IEEE, 2017.
\item [4.] Kingma, Diederik, and Jimmy Ba. "Adam: A method for stochastic optimization." arXiv preprint arXiv:1412.6980 (2014).

\item [5.] Krizhevsky, Alex, Ilya Sutskever, and Geoffrey E. Hinton. "Imagenet classification with deep convolutional neural networks." 
Advances in neural information processing systems. 2012.
\item [6.] Verma, Prateek, and Preeti Rao. "Real-time melodic accompaniment system for Indian music using TMS320C6713." VLSI Design (VLSID), 2012 25th International Conference on. IEEE, 2012.
\item [7.] Griffin, Daniel, and Jae Lim. "Signal estimation from modified short-time Fourier transform." IEEE Transactions on Acoustics, Speech, and Signal Processing 32.2 (1984): 236-243.

\end{enumerate}


\end{document}